# Ultra fast bit addressing in a magnetic memory matrix with crossed wire write line geometry

*H.W. Schumacher*

Physikalisch-Technische Bundesanstalt, Bundesallee 100, D-38116 Braunschweig, Germany.

E-mail: hans.w.schumacher@ptb.de,

phone: +49 (0)531 592 2414,

fax: +49 (0)531 592 2205

## Abstract

An ultra fast bit addressing scheme for magnetic random access memories (MRAM) in a crossed wire geometry is proposed. In the addressing scheme a word of cells is programmed simultaneously by sub nanosecond field pulses making use of the magnetization precession of the free layer. Single spin simulations of the free layer dynamics show that the pulse parameters for programming an arbitrary word of the array can be chosen such that the magnetization of the cells to be written performs either a half or a full precessional turn during application of the programming pulse depending on the initial and final magnetization orientation of the addressed cells. Such bit addressing scheme leads to a suppression of the magnetization ringing in all cells of the memory array thereby allowing ultra high MRAM write clock rates above 1 GHz.

# I. INTRODUCTION

Magnetic random access memories (MRAM) (1,2) are now sometimes also referred to as "universal memories" (3) as they combine features which were previously thought as being incompatible like non-volatility, low power consumption, unlimited write endurance and high speed of operation. Concerning the speed of a write operation of an MRAM one limiting aspect is the minimum duration of the magnetic field pulse resulting in reliable magnetization reversal of the magnetic memory cell. A further aspect is related to the damping of magnetic excitations after the decay of the magnetic field pulse - the so called ringing of the magnetization. Such ringing can persist for several nanoseconds thereby limiting the MRAM write clock rate to values below a few hundred MHz. Optimum MRAM write clock rate could in principle be achieved by making use of the so-called ballistic magnetization reversal (4,5). During ballistic magnetization reversal a well tailored transverse magnetic field pulse induces exactly a 180° precessional turn of the magnetization. As the field pulse is matched to the half of the magnetic precession period (i.e. to a 180° turn) the magnetization turns directly from the initial to the final reversed orientation without any remaining magnetization ringing after field pulse decay (6,7,8,9). Unfortunately, the suppression of ringing after the decay of the switching field pulse is only one side of the coin. In the most common MRAM architecture the selection of a certain magnetic memory cell for programming is done via two sets of metallization write lines running row and column wise and intersecting at each magnetic memory cell. Bit addressed switching is performed by applying so-called half select field pulses to one row line and one column line at the same time. A half-select pulses is not sufficient to reverse the magnetization of a cell whereas the superposition of the two half-select pulses at the intersection is. Experiments showed that also superposed pulses generated by a crossed wire geometry allow ultra fast quasi ballistic magnetization switching (10,11). But what about the other cells subjected to the half-select pulse only? Here, a strong ringing can still occur which will limit the MRAM clock speed.

In previous work (12) a first solution to the problem of ringing of the half selected cells in an MRAM was proposed. The underlying concept was to find parameters for the half select pulse fulfilling the following two requirements: (i) the half select field pulse alone induces a full precessional turn of the magnetization (13) and (ii) the superposition of two half select field pulses at the addressed magnetic memory cell induces a half precessional turn of the magnetization. Upon pulse decay in case (i) the magnetization is oriented again along its initial orientation and in case (ii) the magnetization is oriented along the reversed orientation. In both cases the magnetization ringing is suppressed as the magnetization is already in or at least very near equilibrium when the half and full select pulses decay. It was shown that parameters for such ringing free bit addressing can be found for weak magnetization damping and for a special device geometry involving hard axis pulses only. The drawback of this device geometry is however that the memory could only be operated in a toggle switch mode. Programming of a cell into a predefined memory state is thus only possible when the initial state of the memory cell is known and a pre-write read-out of the cell must be performed.

In this work the concept of ringing free bit addressing is extended to a standard crossed wire memory cell layout. The bit addressing scheme allows directed programming of a cell into a predefined state regardless of the initial state of the cell. From symmetry considerations it is shown that optimized bit addressing can be obtained most eas-ily when programming a whole row of memory cells ("word") simultaneously. Numerical simulations of the magnetization dynamics of the cells in a single spin approximation shown that it is possible to tailor the pulses such that for all cells of the word to be written the magnetization performs either a half or a full precessional turn during pulse application. This bit addressing scheme results in a suppressed magnetization ringing in all cells of the MRAM array allowing ultra high write clock rates above 1 GHz per word and thus of several GHz per bit.

## II. MRAM ARRAY LAYOUT

Fig. 1 shows a sketch of the MRAM array layout studied in this work. The ellipticc magnetic memory cells are arranged in a matrix of *m* rows and *n* columns (*n,m* are integer numbers). The easy axis of the cells corresponds to the long axis of the ellipsoids. Each cell comprises a magnetic free layer consisting of a single ferromagnetic layer. The digital information of the cells is stored in the orientation of the magnetization M of the free layer as indicated by the arrows in some of the cells of Fig. 1.

Programming of the cells is performed using the conductive wiring consisting of the bit lines (BL1 … BLn) running vertically and the word lines (WL1 … WLm) running horizontally. The bit lines are aligned parallel to the in plane hard axis of the free layer and the word lines are aligned parallel to the easy axis of the free layer. A current pulse IBL through one of the bit lines (e.g.: BLν) generates an easy axis magnetic field pulse at all cells of the column ν whereas a current pulse IWL in one of the word lines (e.g.: WLν) generates a hard axis magnetic field pulse at all cells of the row ν. The field pulses generated by the bit lines can have two different easy axis orientations to program the cells into the two digital memory states. For the hard axis pulses generated by the world line unipolar fields are sufficient.

In the bit addressing scheme under consideration a whole row of memory cells of the magnetic memory cell array (also called a "word") is programmed at the same time. In the array shown in Fig. **1** the word ν is written. The word ν corresponds to the cells situated along a word line WLν. To program the word ν of the array of Fig. **1** a word line current pulse IWLν is applied to the word line WLν. This word line current pulse IWLν results in a magnetic field pulse $h_Y$ along the in plane hard axis for all cells situated in the row ν of the array. At the same time bit line current pulses IBL1 - IBLn are applied to all bit lines BL1-BLn. The bit line current pulses generate easy axis field pulses $h_X 1 \ldots h_X n$ at the cells of the corresponding bit lines. The easy axis field pulses from the different bit lines BL have the same amplitude but different orientations. The final orientation of the magnetization M of each bit μ of the word ν to be written is determined by the orientation of the easy axis field $h_X μ$ generated by the corresponding bit line BLμ (i.e. by the polarity of the bit line current pulse IBLμ). The bit sequence of the word ν and the corresponding bit line current orientations are sketched on the bottom part of the figure. As in standard MRAM bit addressing schemes the easy axis fields $±h_X$ alone (created by the bit lines BL) are not sufficient to reverse the magnetization M of the cells situated at the corresponding bit lines. Only the combination of the easy axis bit line field $±h_X$ with the hard axis word line field $h_Y$ allows reversal of the cell magnetization. After application of the combined easy and hard axis field pulse $±h_X$ and $h_Y$ the magnetization of each cell combined easy and hard axis field pulse $\pm h_X$ and $h_Y$ the magnetization of each cell will then be aligned along the applied easy axis field $\pm h_X$ of the corresponding bit line BLµ.

During the programming of a word the different cells of the array are subject to different magnetic field pulses. The cells can be classified in four categories **a-d** of different relative orientations of the initial magnetization M of the free layer and of the applied magnetic field H. Examples of cells of the four categories are marked by the circles **a-d** in Fig. **1**. The relative field orientations of the four categories are sketched in more detail in Fig. **2 a-d**. The cells which do not belong to the word ν to be written are only subject to the easy axis field generated by the bit lines BL. Depending on the initial easy axis orientation of the magnetization M of the cells and on the orientation of the bit line easy axis field $\pm h_X$ the applied easy axis field is either oriented parallel to the initial magnetization as in case **a** or antiparallel as in case **b**. In both cases the final orientation of the magnetization after pulse application should be the same as the initial one (see Fig. **2 a,b** on the right hand side). Also in the programmed word ν two different relative orientations of the field vector H = ($\pm h_X, h_Y$) and the magnetization M can occur. For all cells in the word ν the applied field is tilted out of the easy axis due to the non vanishing hard axis field component $h_Y$ generated by the word line WLν (Fig. **2 c,d**). Again one can distinguish between the case **c** where the easy axis component $h_X$ is oriented parallel to the initial magnetization M and the case **d** where the easy axis component $h_X$ is oriented antiparallel to M. Also in the case **c** (like in the cases **a** and **b**) the applied field pulse should not induce magnetization reversal and the initial and final magnetization states are the same. Only in case **d** the magnetization of the cell must be reversed after the application of the magnetic field pulse. In the four cases only switching from the initial negative easy axis orientation to positive orientation is considered. Due to the symmetry the reverse switching process (i.e. from positive to negative easy axis orientation) is of course also covered by the four cases of field application. Note again, that these four cases cover all possible relative orientations of initial magnetization and applied field in this bit addressing scheme. Either the cells are subject to an easy axis half select field $H = (\pm h_X, 0)$ generated by the bit lines (cases a,b) or the cells are subject to a tilted full select field $H = (\pm h_X, h_Y \neq 0)$ generated by the bit and word line. As the whole word is programmed simultaneously a further case of hard axis half selection generated by the word line does not occur.

As already mentioned the aim is to reduce the magnetization ringing in the array of cells after the decay of the bit addressing pulses. In this bit addressing scheme the ringing is most pronounced in the programmed word v as here the largest tilt of M out of the easy axis during pulse application occurs. This ringing could be suppressed when the following conditions are met:

(i)     when M is not to be reversed upon pulse decay (case c) M undergoes a *full* precessional turn about the effective field during field pulse application.

(ii)    when M is to be reversed upon pulse decay (case d) M undergoes a *half* precessional turn about the effective field during field pulse application and performs a ringing free ballistic switching trajectory (5).

In both cases the free layer magnetization M of all cells of the word to be written is again oriented very near the easy axis and thus near the equilibrium conditions. Conse-quently, the ringing of the magnetization M of the free layer will be strongly reduced if not fully suppressed.

Such suppression of ringing after a full precessional turn (no switch) (13) and after a half precessional turn (ballistic switch) (7) has been achieved in microscopic magnetic memory cells by tailoring the duration and amplitude of the applied magnetic field pulses. One thus has to properly tailor the duration $T_{Pulse}$ and the field values for the pulse $H_S$ that switches the magnetization and for the pulse $H_{NS}$ that does not switch the magnetization under the constraints of the given device geometry. The pulse parameters then have to fulfill the "no-ringing criterion" given by

$$T_{pulse} = T_{Prec}(H_{NS}) = \tfrac{1}{2}\, T_{Prec}(H_S).$$

Here, $T_{Prec}(H_{NS})$ is the duration of the first precessional turn upon application of the "no-switch" field $H_{NS}$ and $\tfrac{1}{2} T_{Prec}(H_S)$ is the duration of the first half precessional turn upon application of the switching field $H_S$. If one is able to find field values $H_S$, $H_{NS}$ fulfilling the right hand side part of the no ringing criterion one just has to choose the pulse duration $T_{pulse}$ accordingly. Then ultra fast and practically ringing free MRAM write operation is possible.

### III. Optimization of the Bit Addressing Pulse Parameters

To obtain this optimal situation in our device geometry the parameters of the easy axis and hard axis field pulses generated by the bit and word lines have to be adapted to the specific magnetic properties of the free layer of the cells. Parameters of the field pulses which can be varied are e.g. the easy and hard axis field components, the pulse duration $T_{Pulse}$, and the rise and fall times of the easy axis and hard axis field pulses $T_{rise}$, $T_{fall}$. The properties of the free layer which play a role are e.g.: the saturation magnetization $M_S$ (which is depends on the magnetic material of the free layer), the geometry of the free layer (and thus the shape anisotropy), the other material dependent anisotropies of the magnetic free layer. Of further importance are also the magnetic coupling of the free layer to the pinned layer and to other layers of the cell stack.

In the following an example is presented of an optimization of the pulse parameters for a certain set of cell parameters and for the device geometry of Fig. **1**. The parameter optimization is done using numerical simulations of the magnetization dynamics of the cell's free layer in a simple single spin model. In this example only the pulse duration $T_{Pulse}$ and the pulse amplitudes along the hard and easy axis are varied to achieve optimum bit addressing trajectories. The free layer of the magnetic cells is modeled as a rotational ellipsoid of permalloy (NiFe) having a saturation magnetization of $4\pi M_S = 10800$ Oe. The demagnetizing factors are $N_X/4\pi = 0.00615$ (easy axis), $N_Y/4\pi = 0.01746$ (in-plane hard axis), and $N_Z/4\pi = 0.9764$ (out of plane) corresponding to ellipsoid dimensions of 500 nm x 200 nm x 5 nm (14). Other anisotropies than shape anisotropy are not taken into account. In the calculations the easy axis is along the x-axis, the intermediate axis (in plane hard axis) is along the y-axis, and hard axis (out of plane hard axis) is along the z-axis as marked in Figs 1 and 2.

The time evolution of the magnetization response to the transverse field pulses is derived by numerically solving the Landau-Lifshitz-Gilbert (LLG) equation (15)

$$\frac{dM}{dt} = -\gamma(M \times H_{eff}) + \frac{\alpha}{M_S}\left(M \times \frac{dM}{dt}\right)$$

in the single spin approximation i.e. under the assumption of a homogeneous orientation of the magnetization in the free layer (4). In the LLG equation $\gamma$ is the gyromagnetic ratio, $H_{eff}$ the effective field, and $\alpha$ is the Gilbert damping parameter. In the simulations a Gilbert damping parameter of $\alpha = 0.03$ is used which has also been measured in realistic magnetic memory cells (7). Due to the symmetry of the system only the trajectories for switching from negative easy axis orientation to positive easy axis orientation is computed. Initially, i.e. before pulse application M is practically oriented along the negative easy axis orientation. Only a small initial tilt of M of 1° in plane out of the easy axis orientation is assumed: $M = (m_X, m_Y, m_Z) = (-0.98, 0.02, 0) \approx (-1,0,0)$. The hard axis field component is always applied in the $+h_Y$ direction. The easy axis field component is either applied in the negative easy axis direction ($-h_X$, no switch, case c) or in the positive easy axis direction ($+h_X$, switch, case d).

During bit addressing the rise of the tilted field pulse H induces a precessional motion of the free layer magnetization M. M then precesses about the vector of the effective field $H_{eff}$ being the sum of the applied field $H = (h_X, h_Y, 0)$ and an internal field $H_{int}$. In the general case $H_{int}$ is due to the various magnetic anisotropies of the free layer and comprises the demagnetizing field $H_D$. The internal field $H_{int}$ depends on the relative orientation of H and M. Consequently, also the precession period $T_{Prec}$ (H) depends on the relative orientation of the applied field H and the initial orientation of M.

This dependence of the computed precession period on the field strength and field orientation is displayed in Fig. 3. Fig. 3 (a) shows the duration of the first full precession period $T_{Prec}(H)$ upon application of a magnetic field step H. Fig. 3 (b) shows the duration of the first *half* precessional turn $½T_{Prec}(H)$ upon application of the same magnetic field step H. The values of $T_{Prec}$ and $½T_{Prec}$ are derived from the LLG simulation of the magnetization response to an in-plane field step having an amplitude $H=(h_X,h_Y,0)$ and a realistic rise time of $T_{rise} = 100$ ps. In Fig. 3 (a,b) the values of the (half) precession period are grey scale encoded. The grey scale map is plotted as a function of the hard axis field component $h_Y$ (horizontal axis) and of the easy axis field component $h_X$ (vertical axis). Long durations of the precession period are encoded in dark and short durations are encoded light (see grey scale bar on the right hand side). The scales and grey scale encodings of the two maps in (a) and (b) are identical. For symmetry reasons only data for positive hard axis field values is computed and displayed ($h_Y > 0$).

As already mentioned the initial magnetization M is oriented along the negative easy axis orientation M ≈ (-1,0,0). In the Figure M thus initially points to the bottom. In the upper half of the plots M is initially oriented antiparallel to the easy axis field component as shown in the sketch in (b). In the lower half of the plots (i.e. for negative easy axis fields -$h_X$) M is initially oriented parallel to the easy axis field component -$h_X$ as shown in the sketch in (a). In parts (a) and (b) of Fig. 3 light values corresponding to short (half) precession periods are found in the lower half i.e. for parallel easy axis field components. The applied field H and the internal field $H_{int}$ are adding up resulting in a strong effective field $H_{eff}$ and thus in short precession periods. The longest (half) precession periods are found in the dark region running from the upper left to the centre right of both plots. In this region the transition from non-switching to switching occurs. Below this transition the magnetization does not overcome the hard axis during pulse application and no magnetization reversal is possible. Above this region M overcomes the hard axis during pulse application and magnetization reversal occurs. The transition corresponds approximately to the well known Stoner-Diamond describing the quasi static magnetization switching thresholds. Near this switching limit the applied field H and the internal fields $H_{int}$ are approximately of the same strength but of opposite sign. The effective field $H_{eff}$ thus nearly vanishes resulting in very long precession periods. Above this limit the applied fields starts to dominate and the effective field values rise again resulting in a decrease of the precession periods with increasing applied field amplitude.

Let us now consider the consequences for the two relevant cases for MRAM operation. In the case **c** of Fig. 2 the field for non switching $H_{NS}$ has a non-vanishing hard axis component $h_Y \neq 0$ and an easy axis component along the initial magnetization. The easy axis component $-h_X$ is negative. Under the application of $H_{NS}$ the magnetization should perform a full precessional turn. The relevant values are thus found in the lower half of Fig. **3(a)**. Here, the duration of the full precession period $T_{Prec}$ is *short* due to the influence of the parallel internal fields. Conversely, case **d** of Fig. **2** corresponds to a field with the same non vanishing hard axis component $h_Y \neq 0$ and a positive easy axis component $+h_X$. Here M should perform approximately a half precessional turn under the application of the switching field pulse $H_S$. The relevant values are thus found in the upper half of Fig. **3(b)** slightly above the dark region of the switching limit. Here, the duration of the half precession period ½ $T_{Prec}$ is *long* due to the influence of the antiparallel internal fields. The existence of a region of long *half* precession periods for an antiparallel easy axis field orientation and of a region of short *full* precession periods for the opposite easy parallel axis field orientation could allow to find a set of field values allowing optimum bit addressing according to the no-ringing criterion. One thus has to find a pair of field values ($\pm h_X, h_Y$) having the same hard axis component $h_Y$ and an easy axis component of opposite sign and equal absolute value $\pm h_X$. The field value for switching is $H_S$ = ($+h_X, h_Y$) is found in the upper half of Fig. 3(b) and the field value for non-switching $H_{NS}$ = ($-h_X, h_Y$) is found in the lower half of Fig. 3(a).

To make it easier to find the corresponding field values in the data of Fig. **3** a new figure of merit for the no-ringing criterion is computed: the precession mismatch $\Delta T(|h_X|, h_Y)$. The precession mismatch is calculated for each field value pair ($\pm h_X, h_Y$) and is the absolute value of the difference of duration of the first half precession period $\tfrac{1}{2}T_{Prec}(+h_X, h_Y)$ at positive easy axis fields and of the duration of the full precession period $T_{Prec}(-h_X, h_Y)$ at negative easy axis fields: $\Delta T(|h_X|, h_Y) = |\tfrac{1}{2}T_{Prec}(+h_X, h_Y) - T_{Prec}(-h_X, h_Y)|$. If $\Delta T = 0$ the durations of the half and the full precession period at the field value pair ($\pm h_X, h_Y$) match and the right hand side of the no-ringing criterion is exactly fulfilled.

In Fig. **4a** the precession mismatch $\Delta T(|h_X|, h_Y)$ is displayed in a grey scale map. $\Delta T$ is plotted as a function of the hard axis field (horizontal axis) and of the absolute easy axis field $|h_X|$ (vertical axis). In the black region marked by the arrow the precession mismatch is smaller than 10 ps. Here, the right hand side equation of the no-ringing criterion is approximately fulfilled which is a prerequisite for optimized bit addressing. The second prerequisite is that the switch field value $H_S$ results in ballistic magnetization reversal and that the ringing upon switch pulse application is suppressed. In Fig. **4b** the switching field threshold and the field values for which ballistic reversal trajectories are possible are displayed. As mentioned before for field values above the switching field threshold the magnetization trajectories overcome the hard axis during field step application. The field region above the switching field threshold is marked in grey. Furthermore, the fields for which no-ringing trajectories are possible (5) are marked in black. Here, the magnetization trajectories pass through the reversed equilibrium position $M \approx (1,0,0)$ during the application of the field steps. When the pulse decay occurs at this point of the trajectory optimum ballistic magnetization reversal occurs. For optimized bit addressing the field value $(\pm h_X, h_Y)$ should thus also be chosen from the black region of the no-ringing switching trajectories.

In Fig. **5** the relevant parts of Fig. **4a** and **b** are overlaid in the same plot. The region of small precession mismatch $\Delta T < 10$ ps and the region of no-ringing switching are differently hatched. The two regions overlap at the cross hatched a field range around $|h_X| \approx 26\text{-}36$ Oe and $h_Y \approx 77\text{-}81$ Oe marked by the arrow. In this field range optimized bit addressed magnetization reversal in the array is possible if the pulse duration $T_{pulse}$ of the easy and hard axis pulses is properly matched to the duration of the precession periods according to no-ringing criterion.

## IV. BIT ADDRESSING TRAJECTORIES

In the following a set of magnetization trajectories which allow ultra fast MRAM operation is presented. The easy and hard axis field values are taken from the region of optimum bit addressing of Fig. 5. The time evolution of the hard and easy axis field pulses and the time dependence of the magnetization trajectories upon application of these field pulses is displayed in Fig. 6. A set of optimized bit addressing pulses taken from the cross hatched field range of Fig. 5 is shown in the uppermost panel. The pulse duration of the pulses is matched to the corresponding precession periods of the no-ringing criterion. The pulse duration of the hard and easy axis pulse is $T_{Pulse}$ = 325 ps when measured at half of the maximum amplitude. The pulse starts to rise at 0 ps and fully decays to zero at a time of 425 ps. The fall and rise times of the pulse are $T_{rise}$ = $T_{fall}$ = 100 ps. The amplitude of the hard axis field pulse is $h_Y$ = 78 Oe. The amplitude of the easy axis field pulse is $h_X$ = ± 28 Oe. In the four lower panels **a-d** of Fig. **6** the components of the magnetization M = ($m_X$,$m_Y$,$m_Z$) are plotted as a function of time. The relative field orientation in the four cases **a-d** is again sketched on the right hand side. The resulting magnetization trajectories during application of these field pulses are now briefly discussed.

In case **a** only the easy axis field $h_X$ generated by the bit line is acting on the magnetization M of the free layer. The magnetic field pulse H = (-28 Oe,0) is applied parallel to the initial magnetization. Consequently, the torque on the magnetization M according to the LLG equation practically vanishes and no precession of the magnetization occurs. Furthermore the magnetization is not reversed upon field pulse decay ($m_X$ = -1). Also in case **b** only an easy axis bit line field pulse H = (+28 Oe,0) is acting on the magnetization M. Now the field is applied antiparallel to the initial magnetization. Also in this case the torque on the magnetization according to the LLG equation is practically zero. Furthermore the applied easy axis field is well below the easy axis switch field threshold and practically no tilt of the magnetization occurs. Consequently, no magnetization ringing is found and the magnetization is not reversed upon field pulse decay ($m_X = -1$).

In case **c** the applied field components are $h_X = -28$ Oe and $h_Y = 78$ Oe. The easy axis component is oriented parallel to the initial magnetization. Here the magnetization performs approximately a full precessional turn about the effective field during pulse application. Due to this full precessional turn the magnetization is oriented near the initial easy axis upon pulse decay and only little ringing of the magnetization after pulse application occurs. The slight remaining ringing is however inevitable due to the finite damping parameter of $\alpha = 0.03$. The damping results in a relaxation of the magnetization towards the tilted effective field during precession. In this example at the moment of pulse decay M is still tilted by about 11° out of the easy axis (see arrow). Due to this rather small tilt the resulting ringing is not very pronounced. After a first precessional "ringing" turn of the magnetization about the internal fields after field pulse decay (i.e. at a time of about 700 ps after the rise of the field pulse) the maximum occurring tilt is less than 6° out of the easy axis. The magnetization ringing can thus be neglected within less than 1 ns after the initial rise of the magnetic field pulse. Furthermore, the magnetization is not reversed upon field pulse decay ($m_X = -1$).

In case **d** the applied field components are $h_X = +28$ Oe and $h_Y = 78$ Oe. The easy axis component is now oriented antiparallel to the initial magnetization. Here the magnetization performs approximately a half precessional turn about the effective field during pulse application. After the half precessional turn the magnetization is now oriented along the reversed easy axis ($m_X = +1$) upon pulse decay and practically no magnetization ringing occurs. Note, that after this ballistic magnetization reversal the ringing is even weaker than in the case **c** as the pulse fields are chosen from the region of the no-ringing trajectories allowing an excellent alignment of M along the reversed easy axis direction upon pulse decay.

As already pointed out these four cases cover all possible relative field orientations and thus all possible magnetization trajectories which occur in the cell array during bit addressing. Consequently, after the programming of a word the ringing is suppressed in *all* cells of the array and the following word could be written immediately after. The achievable write rates are above 1 GHz per *word* and thus of several GHz per bit depending on the number of bits per word.

## V. CONCLUSIONS

This work described a way to find field pulse parameters for optimum bit addressing trajectories that allow ultra fast MRAM write operation. The basic concept for optimum MRAM write operation is that (i) optimum switching is obtained when M performs a half precessional turn (ii) optimum non-switching is obtained when M performs a full precessional turn during magnetic field pulse application. Using simple simulations of the time evolution of the free layer magnetization for a certain set of cell parameters optimized parameters for the bit addressing field pulses were derived. In the proposed bit addressing scheme a whole set of cells ("word") is written simultaneously thereby eliminating the case of half selection by a hard axis field pulse only. As a consequence of this elimination only two relative field orientations (cases c and d) which induce a strong torque on the magnetization (and thus can generate a pronounced ringing of the magnetization) occur during bit addressing. For these two cases it is possible to match the preces-sion periods for switching and non-switching trajectories and thus to obtain optimum bit addressing trajectories.

In the simplified model of the magnetic memory cells only the shape anisotropy of the free layer was taken into account. No further anisotropies or any effects resulting from an inhomogeneous magnetization in the sample were included. However, fast reversal experiments on microscopic memory cells have shown that such simple simulations can well describe the basic physics of ultra fast magnetization motion in MRAM cells (6,7,8) and thus that the results could be indeed transferred to realistic MRAM cells. Nevertheless, as a further step micromagnetic simulations which take into account more realistic device parameters and eventually an experimental test of such optimized bit addressing would be desirable. Further modifications like variation of the pulse shape and of the delay between the bit and word line pulse could be possible and should also be tested for the occurrence of no-ringing bit addressing trajectories.

All numerical values of this work like the maximum write clock rate, pulse duration, pulse amplitude, precession periods etc. depend strongly on the device parameters (geometry and demagnetizing factors, further anisotropies, saturation magnetization, etc.). However, the order of magnitude is valid for a realistic memory cell. Furthermore it should be noted that the value of the Gilbert damping parameter $\alpha$ plays an important role for such ultra fast bit addressing. As pointed out in the literature (5) the position of the no-ringing trajectories (compare Figs. **4, 5)** depends on the value of $\alpha$. For decreasing $\alpha$ the position of the no-ringing trajectories is shifted towards the hard axis and for increasing $\alpha$ towards the easy axis. The optimum bit addressing parameters must be changed accordingly. Note that for very low damping parameters ($\alpha \ll 0.01$) the optimum field orientation is along the hard axis. In this case only a toggle switching and not a directed switching of the memory cells is possible as studied in earlier work (12).

Such optimized bit addressing should not only be fast but also stable. Upon pulse termination of all field pulses the final magnetization is very near the equilibrium conditions and thus near the lowest energy state. Thereby reducing the probability of erroneous writing by half-select pulses e.g. due to thermal activation. In the calculated example the ringing of the magnetization in all cell was on a negligible level within about 700 ps after the initial rise of the bit addressing field pulse. The achievable write rates are therefore above 1 GHz per *word* and consequently of several GHz per bit depending on the number of bits per word. Using this bit addressing scheme ultra fast MRAMs operating with GHz write clock rates thus seems possible.

**FIGURE CAPTIONS:**

**FIGURE 1:**

Sketch of the array of magnetic memory cells and of the conductive bit lines (BL) and word lines (WL) for magnetic field generation. To program the word ν of the array (i.e. the cells along the word line WLν) a word line current pulse IWLν is applied to the word line WLν and simultaneously a set of bit line current pulses IBL1-IBLn is applied to all bit lines BL1-BLn. The polarity of the bit line current pulses IBL defines easy axis field orientation and thus the digital state of the word to be written as sketched on the bottom. Four cases (a-d) of different relative orientation of the generated magnetic field and the initial magnetization are marked by the dotted circles.

**FIGURE 2:**

Detailed sketch of the four cases (a-d) of relative orientation between the applied field H and the initial free layer magnetization M during programming. The initial states are sketched on the left hand side and the final states on the right hand side. (a) H is applied parallel to the M. (b) H is applied antiparallel to M. (c) H is tilted with respect to M. The easy axis component $-h_X$ is parallel to M. (d) H is tilted with respect to M. The easy axis component $+h_X$ is antiparallel to M. Only the case (d) should result in magnetization reversal.

**FIGURE 3:**

Computed duration of the first *full* precessional turn $T_{Prec}(H)$ (a) and of the first *half* precessional turn ½ $T_{Prec}(H)$ (b) upon application of an in plane magnetic field step $H = (\pm h_X, h_Y)$. $T_{Prec}(H)$ and ½$T_{Prec}(H)$ are plotted as a function of the easy axis field component $h_X$ (vertical axis) and of the hard axis field component $h_Y$ (horizontal axis). The rise time of the field step is 100 ps. The magnetization is initially oriented along the negative easy axis $M \approx (-1,0,0)$. The lower (upper) half of the plots corresponds to fields having an (anti)parallel easy axis component as sketched in the pictograms. Light shades correspond to short (half) precession periods and dark shades correspond to long (half) precession periods. The longest values (black) of $T_{Prec}(H)$ and ½ $T_{Prec}(H)$ are found near the transition from switching to non-switching trajectories.

**FIGURE 4:**

(a) grey scale plot of the precession mismatch $\Delta T(|h_X|, h_Y)$ as computed from the data of Fig. 3(a,b) (see text). $\Delta T$ is plotted as a function of the absolute value of the easy axis field component $|h_X|$ (vertical axis) and of the hard axis field component $h_Y$ (horizontal axis). In the black region marked by the arrow the durations of the first full and half precession periods at opposite easy axis fields approximately match. Here, $T_{Prec}(-h_X, h_Y)$ and ½$T_{Prec}(+h_X, h_Y)$ differ by less than 10 ps. (b) plot of the field threshold for magnetization reversal as a function of the easy axis field component $h_X$ (vertical axis) and of the hard axis field component $h_Y$ (horizontal axis). In the white region no magnetization reversal is possible. In the light grey region M overcomes the hard axis during field step applica-tion and magnetization reversal occurs. The black line marks the field values for which ringing free ballistic bit addressing trajectories are possible.

**FIGURE 5:**

Field range allowing optimum bit addressing. The easy axis field component $h_X$ is varied along the vertical axis and the hard axis field component $h_Y$ along the horizontal axis. The region of good precession match $\Delta T(|h_X|,h_Y) < 10$ ps of Fig. 4(a) is displayed together with the region of occurrence of no-ringing trajectories of Fig. 4(b). Both regions are hatched in different directions. The two regions overlap in the cross-hatched region for field values of $|\pm h_X| \approx 26\text{-}36$ Oe and $h_Y \approx 77\text{-}81$ (arrow). For these field values optimum bit addressing parameters when programming a word of the MRAM array can be obtained.

**FIGURE 6:**

Applied field pulses and magnetization trajectories for the four relevant cases (a-d) for optimized bit addressing. The pulse fields are taken from the cross hatched region of Fig. 5. The applied field components are plotted *vs.* time in the uppermost panel. The field amplitudes are $h_X = \pm 28$ Oe and $h_Y = 78$ Oe. The pulse duration is matched to the precession period according to the no-ringing criterion. The rise and fall time of the pulses are 100 ps. The pulses start to rise at the time 0 ps and decay to zero at 405 ps. The pulse duration measured at half of the maximum amplitude is 325 ps. (a-d) Magnetization components $m_X, m_Y, m_Z$ *vs.* time during and after application of the field pulses shown in the upper panel. The relative field orientation in the four cases (a-d) is again sketched on the right hand side. Only the case (d) induces magnetization reversal. Only in the case (c) a slight tilt of the M upon field pulse decay (arrow) and a subsequent ringing of the magnetization occurs. The weak ringing can be practically neglected within less than 1 ns after the initial rise of the bit addressing field pulses. Such bit addressing trajectories allow MRAM write clock rates above 1GHz.

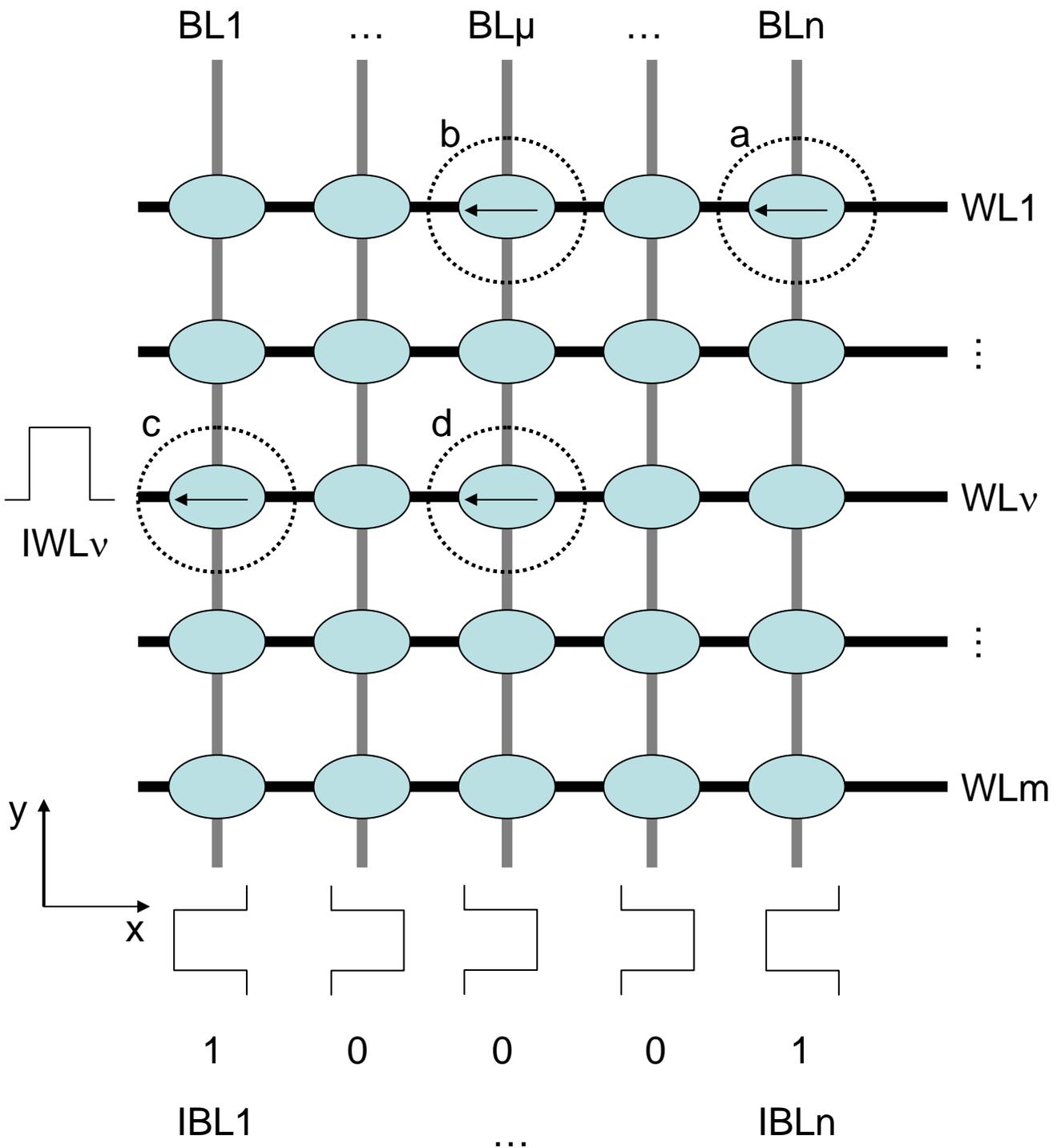

H. W. Schumacher
Figure 1

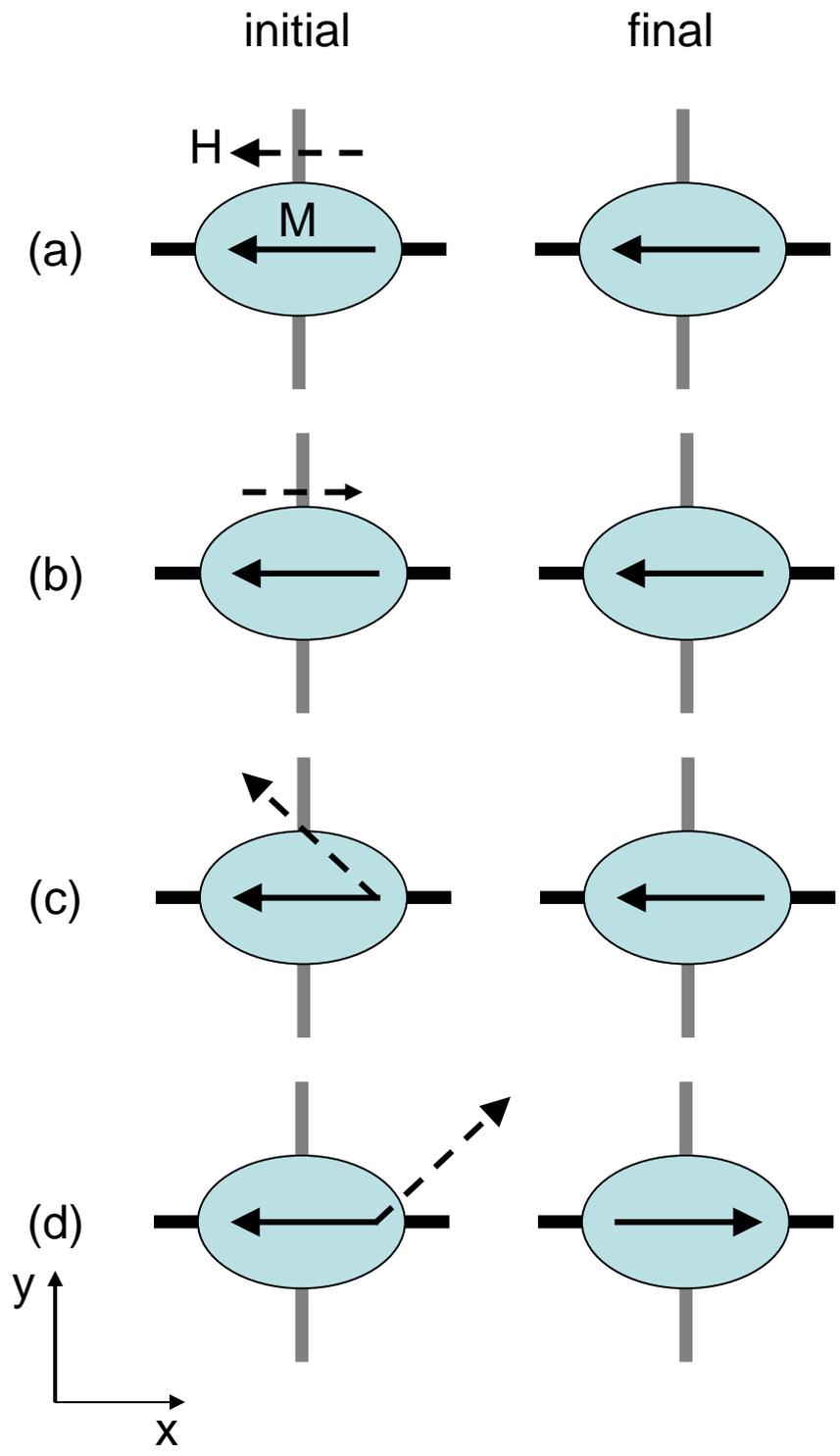

H. W. Schumacher
Figure 2

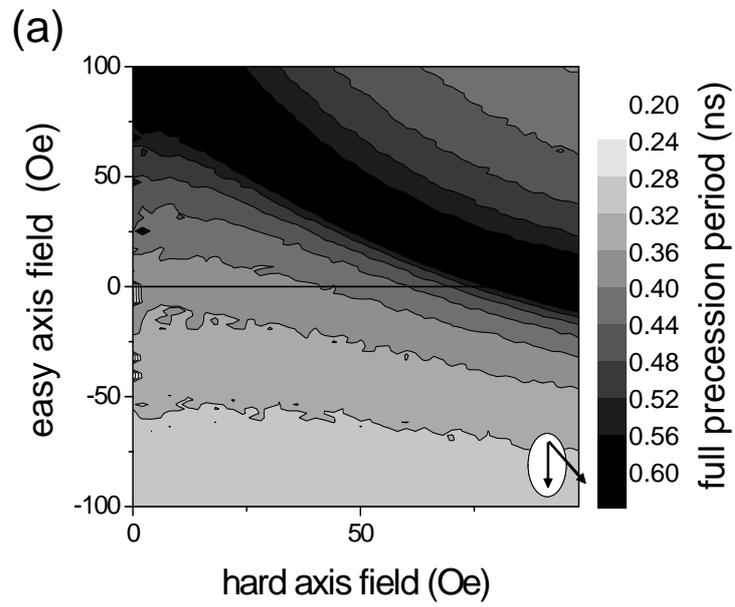

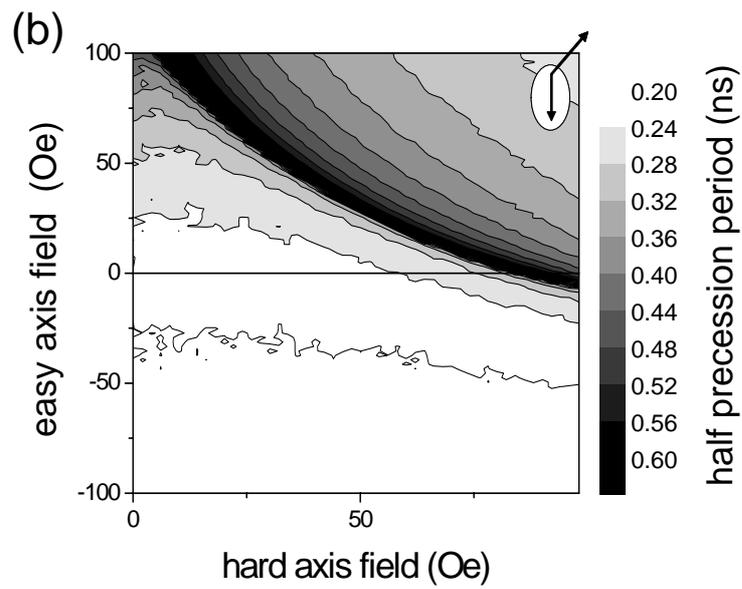

H. W. Schumacher
Figure 3

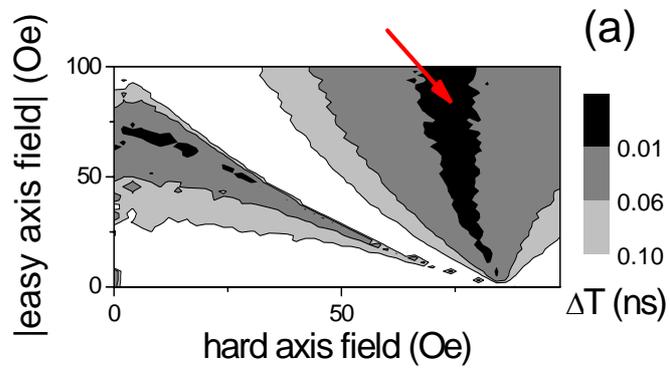

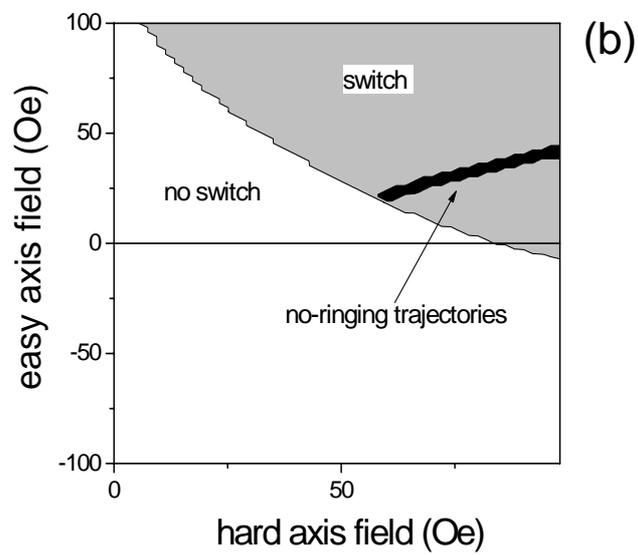

H.W. Schumacher

Figure 4

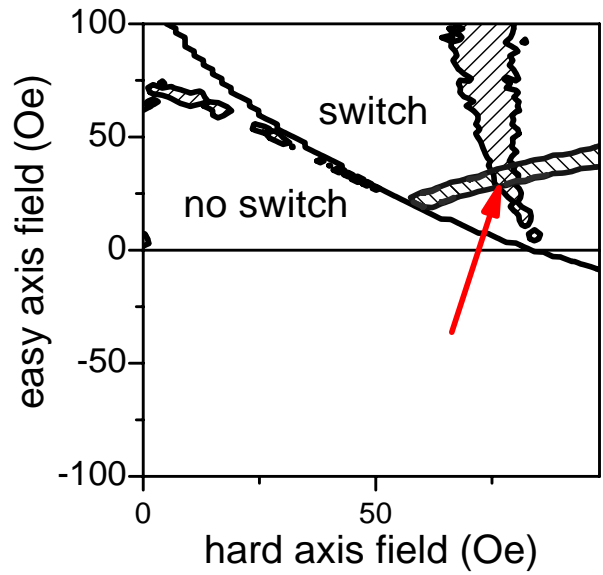

H.W. Schumacher
Figure 5

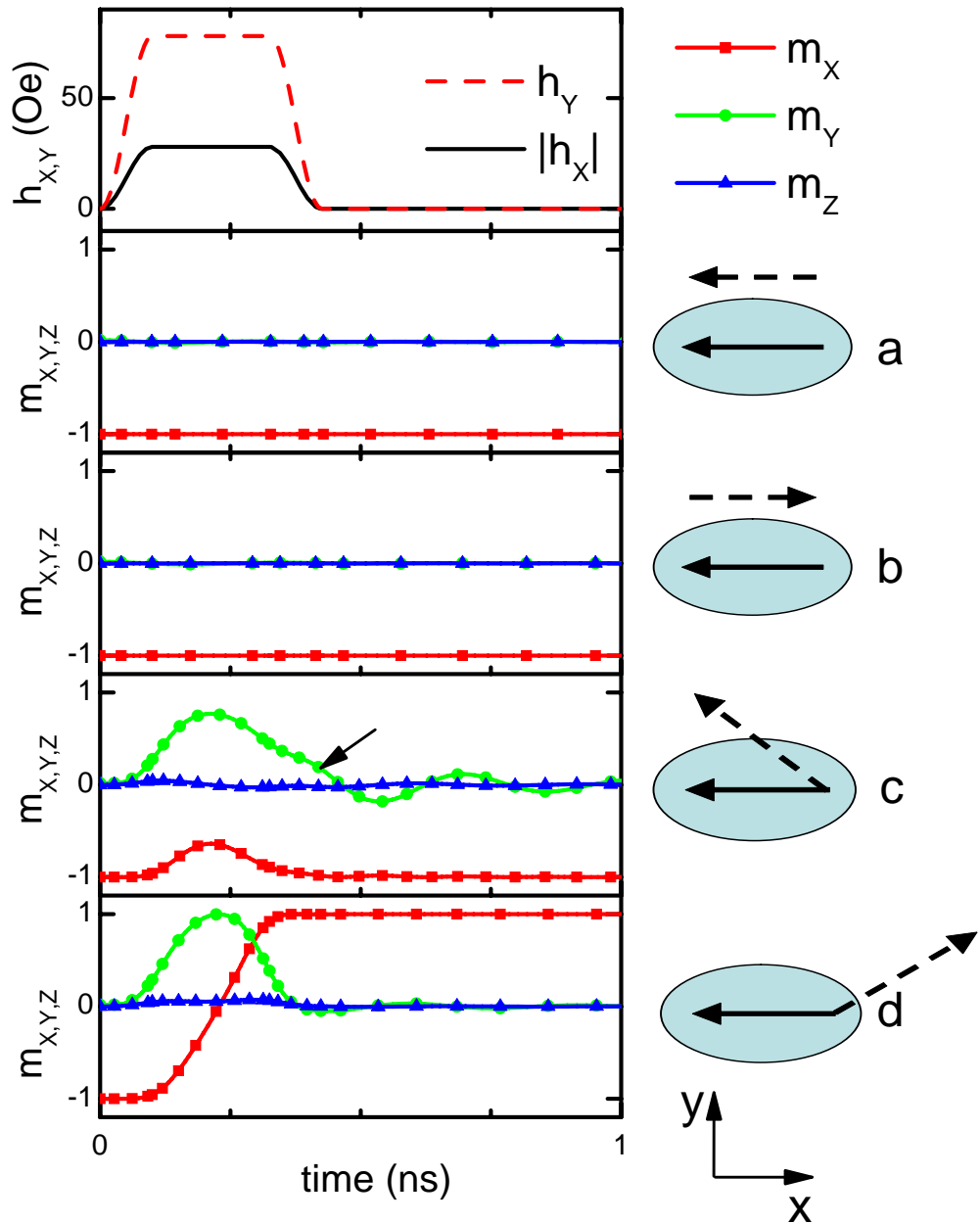

H.W. Schumacher

Figure 6